\documentclass[useAMS,usenatbib,usegraphicx]{mn2e} 
\usepackage{amssymb} 
\usepackage{times} 
\usepackage{aas_macros} 
\usepackage[usenames]{color} 

\usepackage{natbib}
\bibpunct{(}{)}{;}{a}{}{,} 

\usepackage{graphicx}
\usepackage{txfonts}

   \title[Testing excitation in roAp stars]{Testing excitation models of rapidly oscillating Ap stars
     with interferometry}

   \author[M.S. Cunha et al.]
 {M.~S.~Cunha$^{1,2}$, D.~Alentiev$^3$, I.~M.~Brand\~ao$^1$, K.~Perraut$^4$\\
$^1$~Centro de Astrof\'{\i}sica da
Universidade do Porto, Rua das Estrelas, 4150-762 Porto,
Portugal, mcunha@astro.up.pt\\
$^2$ Sydney Institute for Astronomy (SIfA), School of Physics, University of Sydney, NSW 2006, Australia\\
$^3$ Department of Physics, Tavrian National University, Vernadskiy's
Avenue 4, 95007 Simferopol, Ukraine\\
$^4$ UJF-Grenoble 1 / CNRS-INSU, Institut de Plan\'etologie et d'Astrophysique de Grenoble (IPAG) UMR 5274, Grenoble, F-38041, France}

\begin{document} 

\maketitle 
 
\begin{abstract} 
Rapidly oscillating Ap stars are unique objects in the potential
     they offer to study the interplay between a
     number of important physical phenomena, in particular,  pulsations,
     magnetic fields, diffusion, and convection.  Nevertheless, the
     simple understanding of how the observed pulsations are excited
     in these stars is still in progress. In this work we perform a
     test to what is possibly the most widely accepted excitation
     theory for this class of stellar pulsators. The test is based on
     the study of a subset of
    members of this class for which stringent data on the fundamental
     parameters are available thanks to interferometry.  For three out of the four stars
     considered in this study, we find that linear, non-adiabatic
     models with envelope convection suppressed around the magnetic
     poles can reproduce well the frequency region where oscillations
     are observed. For the fourth star in our sample no agreement is
     found, indicating that a new excitation mechanism must be considered. For
     the three stars whose observed frequencies can be explained by
     the excitation models under discussion, we derive the minimum angular extent of the region
    where convection must be suppressed. Finally, we find that the frequency
    regions where modes are expected to be excited in these models
    is very sensitive to the stellar radius. This opens the
     interesting possibility of determining this quantity and related
     ones, such as the effective temperature or luminosity, from
    comparison between model predictions and observations, in other
    targets for which these parameters are not well
    determined. 
\end{abstract} 
 

%

\section{Introduction}

Rapidly oscillating Ap (roAp) stars are chemically 
peculiar stars which pulsate in modes of high radial order \citep{kurtz82} and are
found among the coolest subgroup of Ap stars, usually showing abundance anomalies in lines of ions
of Sr, Cr, rare earth elements and in the core of the hydrogen line. In addition to
specific chemical peculiarities, these cool Ap stars are characterized by their
slow rotation, with periods that range between days and many decades, and strong magnetic fields,  with typical intensities
of a few kG, but which in some stars can reach values higher than 20
kG \citep[{\it e.g.}][]{hubrigetal05}.

Presently there are about 45 roAp stars known with oscillation periods
varying between a little more than 20 minutes
\citep{alentievetal12,elkinetal05} and a little less than 6 minutes
\citep{saioetal12,kreidl86}. There is some observational indication
that the roAp stars with long pulsation periods are relatively evolved,
{\it  i.e.}, that they are approaching the terminal age main sequence. However, the distribution of
roAp stars in the HR diagramme and, hence, their evolutionary status
is hard to establish due to the difficulty in determining their effective
temperatures, which, in turn, results from the surface chemical
peculiarities and the associated horizontal and vertical abundance gradients. To complicate matters further, accurate parallaxes are
available only for about one third of the known roAp stars. 
Despite this, the data available so far is enough to establish that roAp
stars are positioned in the main-sequence part of the classical
instability strip,  having effective temperatures between about 6400
to 8100 K and luminosities between log($L/{\rm L}_\odot)\sim 0.8$ and
log($L/{\rm L}_\odot)\sim 1.5$  \citep[see, {\it e.g.},][for observational and
theoretical reviews on roAp stars]{2009CoAst.159...61K,cunha07}.

The mechanism responsible for the excitation of the oscillations
observed in roAp stars is not yet fully understood. Over the years
there has been a number of suggestions regarding this mechanism,
ranging from the direct effect of the Lorentz force to magnetic
overstability and the opacity mechanism \citep[see][and references
therein]{cunha02}. Nevertheless, only in a few cases  non-adiabatic
calculations leading to the computation of mode growth rates were performed, to
investigate on mode stability. 

Unstable, high radial order modes, similar to those observed in roAp stars, were
found in the models presented by \cite{gautschy98}, \cite{balmforthetal01},
\cite{saio05}, and \cite{theadoetal09}. These models differ
substantially in their physical assumptions. 
\cite{gautschy98} considered that roAp stars may have a chromosphere
and made an ad-hoc modification to a Temperature-optical depth relation such as to induce a temperature inversion at small optical
depths. \cite{balmforthetal01} considered that the
strong magnetic field present in roAp stars suppresses envelope
convection in some region around the magnetic poles, and, thus, built
a spotted model in which only the magnetic equatorial region
retains envelope convection. In addition, they considered models including
the presence of a polar wind. \cite{saio05} considered models in which envelope
convection is fully suppressed and in which the direct effect of the
magnetic field on the oscillations is taken into account.
 Finally, \cite{theadoetal09}
considered both models with and without envelope convection
suppressed and different metallicity abundances and metallicity
profiles. 
Despite differences in the physical assumptions, in all of these models the excitation of the high radial order oscillations
found as a result of the non-adiabatic computations originates from the opacity mechanism acting in the hydrogen ionization
region.

Even though the computations described above predict the
excitation of high radial order pulsations, not
all observed pulsation properties of roAp stars are well reproduced by
these models. In particular, models do not seem to
predict the instability of the very high frequencies, well above the standard acoustic cutoff, observed in some
roAp stars. In addition, difficulties were
found in reproducing the observed red edge of the instability strip
\citep{cunha02,theadoetal09}. 

The range of frequencies excited by the opacity mechanism depends
strongly on the star's effective temperature and luminosity.  Thus, the comparison between model predictions and observations requires the
accurate determination of these quantities. With the recent
development of interferometers \citep[see][for a review]{cunhaetal07}, the determination of effective
temperatures only weakly dependent on atmospheric models became
possible for a few roAp stars. In this paper we will use these interferometrically-based
effective temperature determinations, along with the stars' luminosities derived from the
combination of Hipparcos parallaxes and the stars' bolometric fluxes,
to test the models proposed by \cite{balmforthetal01}. In Section \ref{model} we will briefly describe
the models and the observables for the four roAp stars that have
measured angular diameters. In Section \ref{results} we will present
the results of the model computations for the four stars and in
Section \ref{discussion} we will discuss the results and conclude.

\section{Models and observables}
\label{model}

\subsection{Observables}
The angular diameters of four roAp stars, namely, $\alpha$~Cir
\citep{brunttetal08}, $\beta$~CrB \citep{brunttetal10}, 10~Aql
(Perraut et al., in press - arXiv:1309.4423)
and
$\gamma$~Equ \citep{perrautetal11}, have recently been determined
using the Sydney University Stellar Interferometer (SUSI), in the
first case, and the CHARA interferometric array, in the latter three
cases. Moreover,  bolometric fluxes were
also calculated for these stars, based on calibrated spectra.
These angular diameters were then used, in combination with the
bolometric fluxes, to derive the stars' effective
temperatures in a way that is less dependent on the
modelling of the complex atmospheric structure of these stars,
than other traditional approaches, such as those based on the analysis
of photometric or
spectroscopic data.  Moreover, Hipparcos parallaxes were combined with
the bolometric fluxes to derive the stars' luminosities. 

The effective temperatures and luminosities found in the works mentioned
above will be used as input for the models presented here. For the
remaining of this paper we shall
refer to these inputs as the {\em interferometric inputs}. These are
listed in Table~\ref{data_global} together with photometric and spectroscopic
temperature determinations published in the literature for the same stars. Amongst the
classical methods used to determine effective temperatures, those based on high resolution
spectroscopy are commonly thought to yield the most reliable results
for chemically peculiar stars. Advanced
atmospheric models, based on the LLMODELS code \citep{shulyak2004},
which include individual abundances and an empirical chemical stratification of
elements,
have recently been used to derive fundamental parameters of a number of roAp
stars, including those under consideration here. We shall refer to these
as the  {\em spectroscopic inputs}. Inspection of
Table~\ref{data_global}  shows that for three out of the four stars in
our sample, the effective temperatures derived from
these models are in reasonable agreement with the values derived from
interferometry. For comparison, we consider also the effective temperatures determined by
\cite{Kochukhovandbagnulo06} based on photometric data. Our choice for the latter, over
other photometric temperature determinations published in the literature, is based
on the fact that the above-mentioned authors computed effective temperatures for
the four stars under consideration through exactly the same procedure.
Hence, this set of photometric values may be considered, as much as
possible, as an homogenous set.  For the remaining of the paper we shall
refer to the values of $T_{\rm eff}$ and $L$ derived by
\cite{Kochukhovandbagnulo06} as the  {\em photometric inputs}. 

In addition to the effective temperatures and luminosities, the test
of the proposed non-adiabatic computations requires knowledge of the
pulsation properties of the roAp stars, in particular of the
characteristic frequencies of the observed modes. These are presented in
Table~\ref{data_seismic}, for the four stars under consideration. For
the present work the actual values of the individual frequencies are
not of importance. Only the range of observed frequencies in
each star needs to be considered, as an indicator of the frequency
region in which modes are excited.  We note however, that recent
observations of roAp stars by the NASA {\it Kepler} satellite
\citep{balonaetal11a,balonaetal11b,kurtzetal11}, as well as the earlier 
debate about the roAp nature of $\beta$~CrB \citep{hatzes04,kurtzetal07},
show that the amplitudes of the oscillations in roAp stars may
in many cases be below observational detection. Moreover, it is clear
from the Kepler data that the amplitudes of the modes observed in a
single roAp star can differ by as much as one order of magnitude. Thus, for the purpose of testing theoretical models, each
frequency range given in Table~\ref{data_seismic} should be taken as
the smallest range possible, {\it i.e. }, to be in agreement with the
observations, theoretical models should predict mode excitation in the
given frequency range, but not necessarily only within that frequency
range.

\begin{table*}
\begin{center}
\caption{Effective temperatures and luminosities for the four test
  stars considered in this work.}
\label{data_global}
\begin{minipage}{0.75\textwidth}
\begin{tabular}{|c|c|c|c|c|c|c|c|}
\hline
\multicolumn{2}{|c|}{\textbf{Star}}&\multicolumn{2}{|c|}{\textbf{Interferometric inputs}} &
  \multicolumn{2}{|c|}{\textbf{Spectroscopic inputs}} &
    \multicolumn{2}{|c|}{\textbf{Photometric inputs}}\\
\hline
\hline

HD & Other &$^{a)}$$T_{\rm eff}$ (K)  & $^{a)}$$L$/L$_\odot$ &
$^{b)}$$T_{\rm
  eff}$ (K) &$^{b)}$$L$/L$_\odot$& $^{c)}$$T_{\rm eff}$ (K) & $^{c)}$$L$/L$_\odot$\\
\hline
\hline

201601 & $\gamma$~Equ &$^1$7364 $\pm$ 235 & 12.8
$\pm$ 1.4 & 7550 $\pm$ 50 & 12.6 $\pm$ 0.9 & 7621 $\pm$ 200 & 12.6 $\pm$ 0.9\\
\hline
176232 & 10~Aql & $^2$7900 $\pm$ 200 & $^2$18.5 $\pm$ 1.6 & 7550 $\pm$ 50 &
$^{3}$18.7 $\pm$ 0.9 & 7925 $\pm$ 200 & 20.9 $\pm$ 2.0\\
 \hline
137909 & $\beta$~CrB & 7980 $\pm$ 180 & 25.3 $\pm$
2.9 &8100 $\pm$ 50 &   23.7 $\pm$ 1.9 & 7430 $\pm$ 200 &  27.5 $\pm$ 1.3
\\
\hline
128898 & $\alpha$~Cir & 7420 $\pm$ 170 &  10.5
$\pm$ 0.6  & 7500 $\pm$ 130 & 10.7 $\pm$ 0.3 & 7673 $\pm$ 200 & 11.0 $\pm$ 0.3\\
\hline
\end{tabular}
\begin{flushleft}
Notes: $a)$ $T_{\rm eff}$ values derived based on
  interferometry and $L$ values derived based on Hipparcos parallax
  and bolometric flux
  \citep[][;Perraut et al., in press (arXiv:1309.4423)]{brunttetal08,brunttetal10,perrautetal11}. $b)$ $T_{\rm eff}$
  and $L$ values derived from advanced atmospheric models and high-resolution spectroscopy \citep{kochukhov2009,shulyak2013,nesvacil2013}. $c)$ $T_{\rm eff}$ values
derived based on Geneva photometric indices and $L$ values derived
based on Hipparcos parallax using standard bolometric corrections
\citep{Kochukhovandbagnulo06}. \\ 
$^1$ This value may be in excess by up
to 110 K due to the presence of a close companion \cite[see][for
details]{perrautetal11}. \\
$^2$ This value corresponds to the average of the two values given in
Perraut et al. (in press - arXiv:1309.4423).\\
$^{3}$ The luminosity value adopted here is the average of
the two values presented for He-normal atmospheric models in the work by
\cite{nesvacil2013}, for an effective temperature of 7550 K, and model
calibrations based on space or ground-based data, respectively. 
\end{flushleft}
\end{minipage}
\end{center}
\end{table*}

\begin{table}
\centering
\caption{Observed frequency range of the pulsations exhibited by the four test
  stars considered in this work.}
\label{data_seismic}
\begin{minipage}{0.3\textwidth}
\begin{centering}
\begin{tabular}{|c|c|c|}
\hline
 HD & Other & Frequency range (mHz) \\
\hline
201601 & $\gamma$~Equ & $^{a)}$ 1.31 - 1.42 \\
\hline
176232 & 10~Aql & $^{b)}$ 1.39 - 1.45  \\
\hline
137909 & $\beta$~CrB &  $^{c)}$ 1.03 \\
\hline
128898 & $\alpha$~Cir& $^{d)}$  2.26 - 2.57 \\
\hline
\end{tabular}
\begin{flushleft}
Notes: $a)$~From \cite{gruberbaueretal08}.  $b)$~From
\cite{huberetal08}. $c)$~From
\cite{kurtzetal07} (single frequency). $d)$~From \cite{brunttetal09}.
\end{flushleft}
\end{centering}
\end{minipage}
\end{table}

\subsection{Models and key parameters\label{models}}
To perform linear, non-adiabatic calculations for the four test
stars considered in this work we followed the approach presented by
\cite{balmforthetal01}. A detailed description of the corresponding models and
physical  assumptions can be found in the latter. Here we briefly summarize those aspects that are
important for the present discussion. The computations are based on
two programs. The first generates an equilibrium envelope model,
which does not include the energy generation core of the star, and the
second solves the linearized non-adiabatic pulsation equations for
radial oscillations. 
The model considers that the strong magnetic field present in roAp
stars is capable of suppressing envelope
convection at least in some region around the magnetic poles. Two
options are possible regarding the suppression of convection. The
first suppresses envelope convection at all latitudes,
producing a spherically symmetric equilibrium model. The second
simulates a model in which convection is suppressed only up to some
co-latitude measured from the magnetic poles. In the latter case, two spherically symmetric equilibrium
models are computed, one with envelope convection suppressed,  hereafter {\it
 the polar model}, and the
other with normal convection, treated with a non-local mixing length
prescription described by \cite{gough77}, hereafter {\it the
  equatorial model}. The two equilibrium models
are then matched such as to guarantee that they have the same internal
structure. That is done by adjusting the luminosity and radius of one of
the models until the same temperature, pressure and helium
abundance is obtained in the two models below the convective
envelope. As a result of the matching of the interior, the outer
  layers of the two equilibrium models will differ, and so will their radii and
  luminosities. However, as discussed in \cite{balmforthetal01}, the difference
in luminosity between the two models will generally be smaller than
that induced by line blanketing in a typical Ap star. As a consequence,
it is not possible to derive the extent of the angular region around the
magnetic pole in which envelope convection is suppressed
directly from observations of the stellar surface. 

When envelope convection is not suppressed at
all latitudes, linear non-adiabatic radial eigenfrequencies are
computed separately for each of
the two spherically symmetric equilibrium models. Moreover, the linearized non-adiabatic equations in the
equatorial model are solved using a non-local, time-dependent
treatment of convection \citep[][and references
therein]{balmforth92}. The eigenfrequencies of the two spherically
symmetric models are then used to compute the eigenfrequencies of
the composite model ({\it i.e.} the model composed of the polar and
equatorial regions), through the application of the variational
principle. 

Clearly the procedure described above is oversimplified in a number of
ways, as it neglects the direct effects of the Lorentz force on the equilibrium
structure, as well as on the pulsations. In addition, it presumes that like in
the case of convective motions, the complex dynamics  that would arise
in the outer layers as a consequence of the horizontal hydrostatic unbalance is prevented by the action of the magnetic
field.  Producing a realistic model for these spotted
stars would require a much higher degree of sophistication, which is
beyond the scope of this paper. Instead, our aim is to test the
model as presented by \cite{balmforthetal01}, which has so far been quite successful in explaining the
observations,  against the best data currently available.

\begin{table*}
\begin{center}
\caption{Model parameters of non-adiabatic calculations. Shown are two star identifications,
  model mass, model radius as derived from the interferometric, spectroscopic
  and photometric inputs, respectively, interior hydrogen and helium
  abundances, surface helium abundance, minimum optical depth and
  outer boundary condition in the pulsation code. For the latter three
entries two options have been used. The standard model uses the
options shown in roman font while other models are built by
swapping to the italic font entries, one at the time. }
\label{models}
\begin{minipage}{1.0\textwidth}
\begin{tabular}{|c|c|c|c|c|c|c|c|c|c|c|}
\hline
 HD & Other & Mass (M$_\odot$) & $^1$ R$_{\rm int}$ (R$_\odot$) & R$_{\rm
   spe}$ (R$_\odot$) & R$_{\rm pho}$ (R$_\odot$) & $X_{\rm int}$ &
 $Y_{\rm int}$ & $Y_{\rm sur}$ & $\tau_{\rm min}$ & $BC$ \\
\hline
201601 & $\gamma$~Equ & 1.75 & 2.20 & 2.08 & 2.04 &  0.705 & 0.278 &
{0.01} -- {\it 0.1} & { 3.5x10$^{-5}$} -- {\it 3.5x10$^{-4}$} &
{reflect.}
-- {\it transm.}\\
\hline 
176232 & 10~Aql & 1.95 & 2.31 & 2.53 & 2.43 &  0.705 & 0.278 &
{0.01} -- {\it 0.1} & { 3.5x10$^{-5}$} -- {\it 3.5x10$^{-4}$} &
{reflect.}
-- {\it transm.}\\
\hline
137909 & $\beta$~CrB & 2.05 & 2.64 & 2.48 & 2.67 &  0.705 & 0.278 &
{0.01} -- {\it 0.1} & { 3.5x10$^{-5}$} -- {\it 3.5x10$^{-4}$} &
{reflect.}
-- {\it transm.} \\
\hline
128898 & $\alpha$~Cir& 1.70 & 1.96 & 1.94 & 1.88 &  0.705 & 0.278 &
{0.01} -- {\it 0.1} & { 3.5x10$^{-5}$} -- {\it 3.5x10$^{-4}$} &
{reflect.}
-- {\it transm.}\\
\hline
\end{tabular}
\begin{flushleft}
Notes: $^1$ The actual inputs to the code are the effective
temperature and luminosity. These radii are derived from those inputs
and due to approximations in the last digits of $T_{\rm eff}$ and $L$
they may differ from the published interferometric radii in the last decimal place. 
\end{flushleft}
\end{minipage}
\end{center}
\end{table*}

In addition to defining the extent of the angular region around the
magnetic pole in which envelope convection is to be suppressed, and
fixing the parameters associated to the treatment of convection, the
proposed non-adiabatic calculations require the following
relevant information:

 1) The minimum optical depth in the equilibrium model, $\tau_{\rm min}$. In the
 atmosphere, the thermal stratification is derived from a temperature-optical depth
 ($T$-$\tau$) relation fitted to a model atmosphere of Kurucz
 \citep{shibahashi85}. The value of $\tau_{\rm min}$ is a free
 parameter of the model.  For oscillations with frequencies well
 below the acoustic cutoff, the form of the eigenfunctions in the
 propagation cavity should be relatively independent of the value of $\tau_{\rm min}$
adopted. Nevertheless, in the case of the high frequency oscillations
observed in roAp stars, the form of the
 eigenfunctions below the photosphere is more sensitive to the details of the overlaying
 atmosphere. With this in mind, we consider two different
 values of  $\tau_{\rm min}$ in our computations, namely, $\tau_{\rm
   min}$=3.5x10$^{-5}$ and $\tau_{\rm   min}$=3.5x10$^{-4}$ in order
 to inspect the effect of changing this parameter on the results.

2) The helium profile in the equilibrium model. The helium profile in
models with envelope convection suppressed is determined according to the prescription presented
in \cite{balmforthetal01}, without a polar wind (so, with the accumulation
parameter, $A$, set to zero). The profile is then characterized by a single
parameter, $Y_{\rm surf}$, that establishes the helium at the surface. In our calculations we
assume that helium settles efficiently when convection is suppressed,
leading to a very small helium abundance throughout the outermost
layers, in particular where the hydrogen ionization takes place. When
envelope convection is maintained, we assume a homogeneous helium
profile. These assumptions are supported by the results of the work by
\cite{theadoetal05}. To see the impact of the settling efficiency on
the results, we consider two cases, corresponding to $Y_{\rm
  surf}=0.1$ and $Y_{\rm surf}=0.01$

 3) The outer mechanical boundary condition to be applied
at the temperature minimum in the pulsation model.  In some roAp stars the
frequencies of the observed oscillations are greater than the acoustic
cutoff frequency, {\it i.e.}, the frequency above which acoustic waves
will propagate away into the atmosphere of the star and dissipate. This fact has been discussed in the
literature at length. In particular, \cite{sousa08} suggested that the
coupling of the oscillations with the magnetic field in the outer
layers of roAp stars results in part of the original acoustic wave energy being
converted into magnetic wave energy. The authors argue that the
magnetic part of the energy is not dissipated, allowing a fraction of
the wave energy to be retained in each pulsation cycle.  In the
non-adiabatic calculations considered here the direct effect of the
magnetic field on the oscillations, and thus its effect of the
reflection of the modes, is not taken into account. Hence, the 
natural boundary condition to apply at the outer boundary is one that
allows the wave to propagate away if its frequency is above the
acoustic cutoff.  That boundary condition is one of the possible
options of the pulsation code used in this work, and is derived from
the matching of the solutions to a plane-parallel isothermal
atmosphere. We shall call it the {\em transmissive boundary condition}. Despite the above, it is clear from the observations that
a mechanism for the reflection of the highest frequency modes is missing
in this non-magnetic model. In an attempt to account for the effect of
such physical mechanism, \cite{balmforthetal01} considered a second
option for the outer boundary condition which
fully reflects the waves.  We shall call it the {\em reflective
  boundary condition}. The latter corresponds to the low frequency limit of
the boundary condition obtained when matching the solutions onto an
infinite isothermal atmosphere. Similarly,  in this work we consider
the two alternative outer boundary conditions described in detail in \cite{balmforthetal01}.

A summary of the properties of the models used for the present study is presented in Table~\ref{models}.
Our default set of models are calculated with $\tau_{\rm
   min}$~=3.5x10$^{-5}$, $Y_{\rm surf}=0.01$,
 and the reflective boundary condition. Also, all our models were computed assuming interior chemical
abundances of helium and hydrogen of, respectively, $Y_{\rm int}$ = 0.278 and $X$ =
0.705 (similar to the solar helium and hydrogen abundances derived
from \cite{serenelli2010} and \cite{grevesseandnoels93}).  We note
that we have considered the impact of adopting, instead, an interior chemical composition
similar to that proposed by \cite{Asplund2009} for the bulk of
the sun and found that the impact on the results was negligible, when
compared with the other effects under study. 
For the stellar mass, we assumed $M$=1.75~M$_\odot$ for $\gamma$~Equ,
$M$=1.95~M$_\odot$ for 10~Aql,
 $M$=2.05~M$_\odot$ for $\beta$~CrB, and $M$=1.70~M$_\odot$ for
 $\alpha$~Cir.  These values are within $\pm$ 0.05~M$_\odot$ of the
 values derived based on stellar evolution, given the spectroscopic input parameters considered in
 Table~\ref{data_global} and the above mentioned envelope chemical
 abundances. They are also very close to the values derived by
 \cite{Kochukhovandbagnulo06} for the same stars based on the
 photometric inputs.

\section{Results}
\label{results}

Figures 1 to 3 summarize the results of the non-adiabatic calculations obtained for the stars in our
sample. In the calculations the oscillations are assumed to have a
time-dependence of the type $e^{-i\omega t}$ where the angular frequency
$\omega = \omega_{r}$+i $\eta$. Thus, the oscillations are
intrinsically unstable if the growth rates, $\eta$, are greater than zero.  

\subsection{Equatorial models  \label{sec: equatorial}}
\begin{figure*}
  \centering
   \includegraphics[width=14.cm]{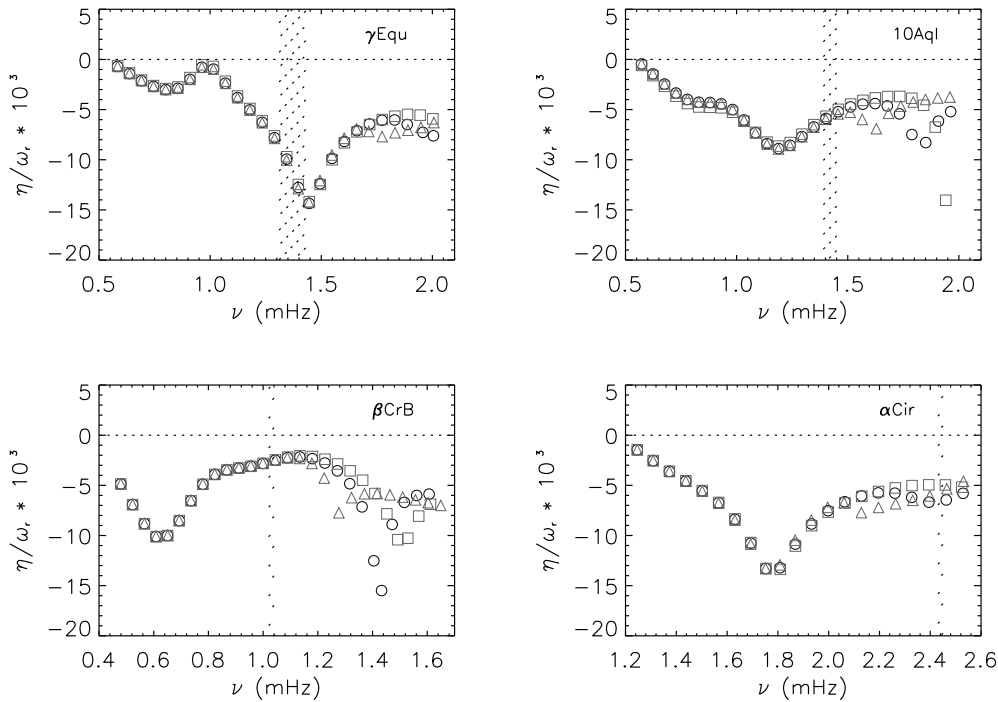}
      \caption{Normalized growth rates for equatorial regions ({\it
          i.e.} without suppression of convection) as function of the
        cyclic frequency $\nu=\omega/2\pi$. Modes are unstable when the
        normalized growth rates are greater than zero. Results are shown
        for the interferometric inputs of: $\gamma$~Equ (top, left),
        10~Aql (top, right), $\beta$~CrB (bottom, left) and $\alpha$~Cir
        (bottom, right).  For each star we show results for the default case
        (circles), for the case
        with $\tau_{\rm min}=3.5x10^{-4}$  (squares) and for the case with
        transmissive boundary condition (upward triangles). All
        equatorial models have homogeneous chemical composition in the
        envelope. In each
        panel, zero
        growth rate is indicated by the horizontal dashed line and the
        shadowed region marks the range of observed frequencies.  }
         \label{equatorial}
   \end{figure*}

The growth rates obtained for equatorial models of the four stars in
our sample are shown in Fig. \ref{equatorial}. For each star, results
are shown for three different models, including the default model and
otherwise similar models but with a different value of the minimum
optical depth or with a different outer
boundary condition in the pulsation code.  The radial orders of the
modes shown are such as to encompass the region of observed frequencies.
These are, respectively, $n$=10-36 in the case of $\gamma$~Equ,
10~Aql, and $\beta$~Crb and $n$=19-39 in the case of $\alpha$~Cir. 

  In all cases, and for all
four stars, modes in the frequency region where oscillations are
observed were found to be intrinsically stable in the equatorial models. From inspection of the gas pressure and
turbulent pressure contributions to the growth rates, obtained from
the computation of the cumulative work integrals \citep[see][for
details]{balmforthetal01},  we found that in all these models the turbulent pressure is
responsible for the stabilization of the high radial order modes, as
found by \cite{balmforthetal01} in their models. From Fig.
\ref{equatorial} it follows also that except for the highest radial orders
(above $n=25$ for 10~Aql and $\beta$~CrB and above  $n=30$, for
$\gamma$~Equ and $\alpha$~Cir), the growth rates are almost independent of the
minimum optical depth adopted in the equilibrium
model, as well as of the outer boundary condition applied in the pulsation code. 

\subsection{Polar models \label{sec:polar}}
\begin{figure*}
   \centering
   \includegraphics[width=14.cm]{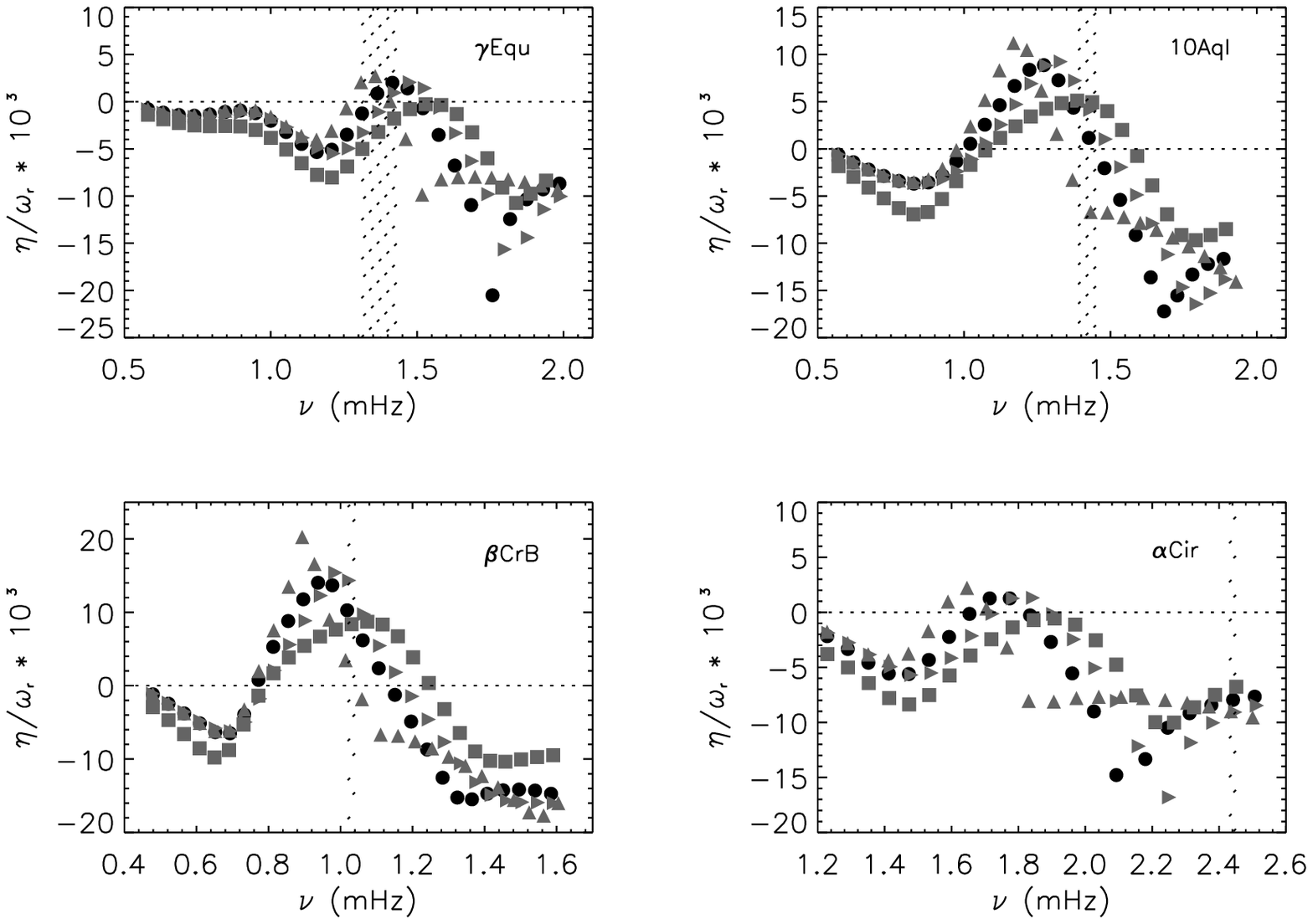}
      \caption{Normalized growth rates for polar regions ({\it
          i.e.} with suppression of convection in the envelope).
       Each panel shows the results for a different star. In addition to the three models described in Fig.
       \ref{equatorial}, a fourth model is considered, with
       $Y_{\rm surf}=0.1$ (rightfacing triangles). Other symbols are used as in Fig.
       \ref{equatorial}, but filled.
              }
         \label{polar}
   \end{figure*}

The growth rates obtained for polar models of the four stars in
our sample are shown in Fig. \ref{polar}. For each star, results
are shown for the same three models and the same ranges of the radial orders
mentioned above.  One additional model, with surface helium abundance
$Y_{\rm surf}=0.1$ is also considered. 

The growth rates for polar
models show a distinct envelope of positive values located
somewhere between radial orders 18 and 28. That envelope is more
prominent in the hotter, more luminous targets, 10~Aql and
$\beta$~CrB, than in the cooler, less luminous targets, $\gamma$~Equ
and $\alpha$~Cir. In fact, for the latter two stars, no positive
growth rates are found in this range of radial orders for models with the
higher value of $\tau_{\rm min}$.  It is also evident from this figure
that the growth rates in polar models depend significantly on the
adopted $\tau_{\rm min}$ and $Y_{\rm surf}$, as well as on the
outer  boundary condition applied in the pulsation code, even at
moderate values of the radial order.

\begin{figure*}
  \centering
   \includegraphics[width=14cm]{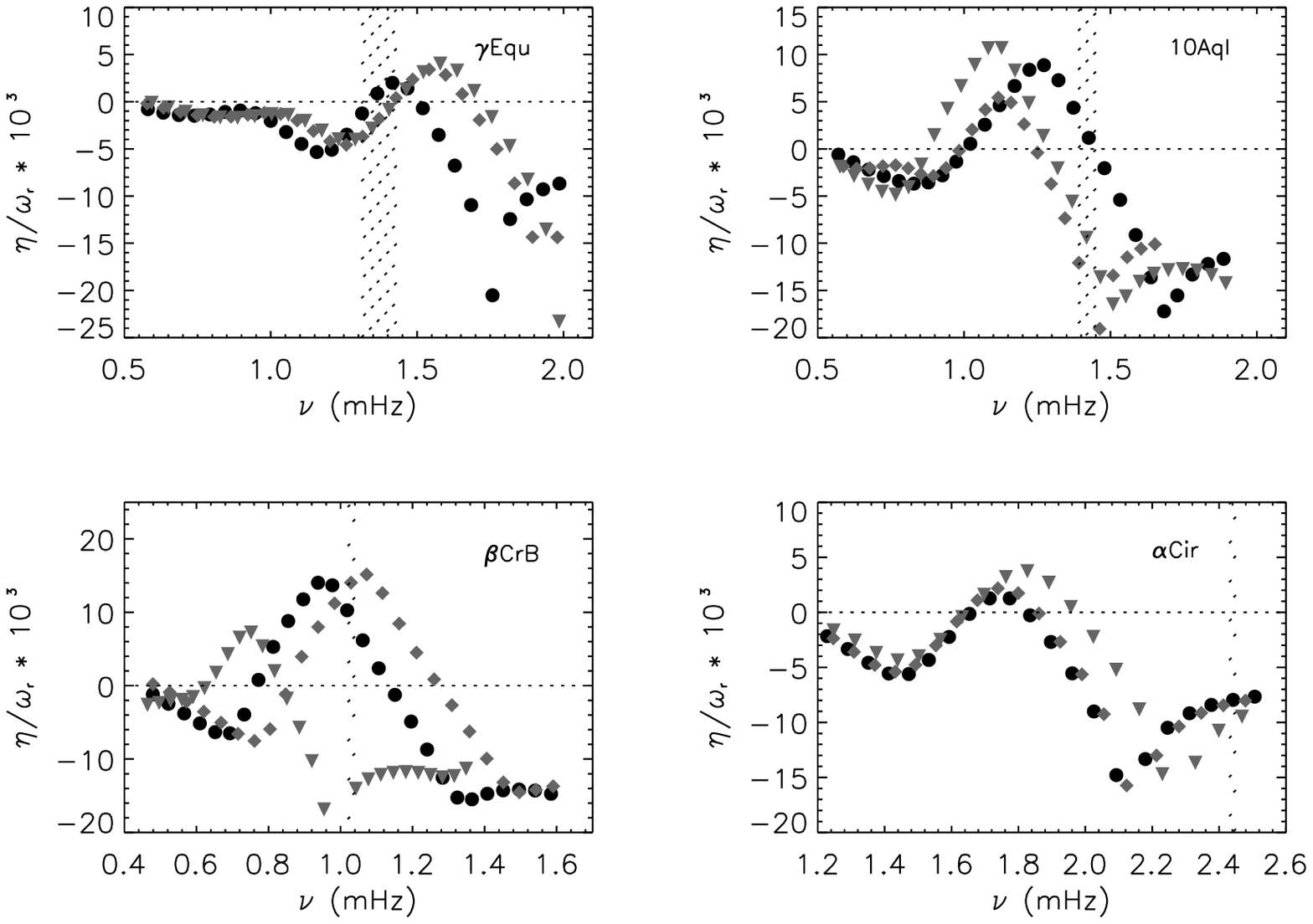}
      \caption{Comparison of normalized growth rates in polar
        regions for different model inputs. Each panel
        shows, for the star under consideration, the results for the default polar
        models computed with the interferometric input
        parameters (circles), the spectroscopic input
        parameters (diamonds), and  the 
       photometric input parameters (downward triangles).
              }
         \label{comp}
   \end{figure*}

The theoretical description for the driving of roAp pulsations under discussion in this work
suggests that the high radial order oscillations
are excited in the polar regions, where convection is suppressed. Thus, we
have computed, for comparison,  default polar models also for the
spectroscopic and photometric input parameters. The comparison between
default models with the different inputs is shown in Fig.
\ref{comp}. For the new models, with photometric and spectroscopic
inputs, we have displayed in the plots the radial orders
overlapping the frequency range adopted for the model with the
interferometric inputs (which do not, in general, coincide with the
radial orders displayed for the latter).  Clearly, both the frequency position of the envelope of positive
growth rates and the magnitude of the growth rates are significantly
affected by the inputs. In particular, it is seen that the position in frequency of the envelope, for
these models of similar mass, reflects essentially the dependence on the
stellar radii through the scaling with $1/R^{3/2}$.

\subsection{Composite models and comparison with observations}

To decide on whether a given mode is excited we need to consider the
growth rates, $\eta^{\rm c}$, in the composite model composed of 
polar  and equatorial regions matched in the way described in Section
\ref{models}. Adopting a spherical coordinate system
(r, $\theta$, $\phi$), and using $\mu=\cos\theta$,  these can be obtained from the expression \citep{balmforthetal01}
\begin{eqnarray}
\eta^{\rm c}_{nlm} & \approx &
\eta_{n0}^{\rm p}\int_{\tilde{\mu}}^1{\left(Y_l^m\right)^2
{\rm d}\mu{\rm d}\phi}+\eta_{n0}^{\rm eq}\int_0^{\tilde{\mu}}{\left(Y_l^m
\right)^2{\rm d}\mu{\rm d}\phi}  \nonumber \\ 
& = & \left(1-\Lambda_l^m\right)\left(\frac{\Lambda_l^m}{1-\Lambda_l^m}
\eta_{n0}^{\rm p}+\eta_{n0}^{\rm eq}\right) .
\label{eq:eta}
\end{eqnarray}
In the above, $\eta^{\rm p}_{n0}$ is the growth rate for the radial
mode of radial order $n$ in the polar model and $\eta^{\rm eq}_{n0}$
is the
growth rate for the same mode in the corresponding equatorial model,
while $Y_l^m$ are spherical harmonic functions, with $l$ and $m$
representing the mode degree and azimuthal
order, respectively. Moreover,  the geometrical factor $\Lambda_l^m$ that accounts for
the extent of the polar region (hereafter, the spot), is given by,
 \[ 
\Lambda_l^m=\left(2l+1\right)\frac{\left(l-m\right)!}{\left(l+m\right)!}
\int_{\tilde{\mu}}^1\left(P_l^m\right)^2{\rm d}\mu , 
\]
where $\tilde{\mu}=\cos\vartheta$ and $\vartheta$ is the angular size of
the region, centred on the magnetic pole, in which convection is
assumed to be suppressed.  We note that equation~(\ref{eq:eta}) assumes that the growth rates of non-radial modes can
be estimated from those of radial modes. As argued by
\cite{balmforthetal01}, that assumption is justified for low degree,
high radial order modes by the fact that the driving takes place in
the outer layers of the star. 

From equation (\ref{eq:eta}) we see that a mode is unstable in the composite model
if [$\Lambda_l^m/(1-\Lambda_l^m)$]$\eta_{n0}^{\rm p}>
-\eta_{n0}^{\rm eq}$. An illustration of the factor
$\Lambda_l^m/(1-\Lambda_l^m)$ for modes of degree up to $l=2$ is presented in figure 4 of
\cite{balmforthetal01}.  With this criterion in mind, next we inspect the growth rates in the equatorial and polar models for
each star in our sample and compare the results with the
observations.  

In the cases of 10~Aql and $\beta$~CrB, the envelope of 
frequencies excited in the polar models quite clearly contains the range of observed
frequencies, almost independently of the conditions assumed for $\tau_{\rm
min}$ and $Y_{\rm surf}$. The match is not as good when the outer
pulsation boundary condition is changed. Nevertheless, we must stress
that no attempt has been made in this work to improve the match of the observations
with the model results, by changing, {\it e.g.} the input parameters
$T_{\rm eff}$, $L$ within their uncertainties, or by changing the
stellar mass and internal chemical abundances. It is quite reasonable
to assume from the frequency-radius scaling discussed in Section \ref{sec:polar} that had
such a search within the input parameter space been performed, good matches of the location of 
observed and model unstable frequencies would have been found for these
two stars also when changing the outer boundary condition. Moreover, a
similar exploitation of the input parameter space could easily lead to
results in which the observed frequencies in these two stars are
centered at the frequency of maximum polar growth rate.
 
From inspection of the growth rates in the equatorial models for 10~Aql we
see that their absolute value in the frequency region of interest is
comparable to the maximum of the growth rates in the excitation envelops
for this star  ($\eta /\omega_{\rm r}\sim$ 5-10 $\times 10^3$) 
in the same frequency region. Taking into
consideration the term $\Lambda_l^m/(1-\Lambda_l^m)$ \citep[{\it cf.} figure 4 of][]{balmforthetal01} we may then expect that excitation will take place
in these models in a frequency region similar to that where modes are
observed if the angular extent of the spot where convection is
suppressed, $\vartheta$, is greater than $\sim$ 30$^\circ$. 
We find from the same reference that modes of  $l$=1 and 2
and $m=0$ are the most easily excited for spots of this size, while
excitation of modes of $l=0$ requires, in this case, $\vartheta >$
60$^\circ$.

In the case of $\beta$~CrB, inspection of the growth rates in the
equatorial models shows that their absolute value in the frequency
region of interest is significantly smaller that the growth rates
found in polar models for the same star and frequency region. The ratio between the latter
and the former is about 5. Evaluation of the term
$\Lambda_l^m/(1-\Lambda_l^m)$ then leads us to conclude that axisymmetric
modes of $l=$ 1 and 2 may be excited in the frequency region of
interest for $\vartheta$ greater that about 20$^\circ$, while radial modes start
being excited if the spot extends more that about 35$^\circ$ in co-latitude. 

We now turn to the case of $\gamma$~Equ. 
Keeping in mind the possibility of adjusting slightly the input parameters
within their observational uncertainties to improve the match between
model results and observations, as discussed above,  we may conclude
that mode excitation is expected in the frequency region of interest in all
polar models of  $\gamma$~Equ considered in this work, except for that with a
larger value of $\tau_{\rm min}$.  Nevertheless, when comparing
results in the polar and equatorial models of $\gamma$~Equ, a very
different picture from that seen in the cases of 10~Aql and $\beta$~CrB
emerges. In the present case, the absolute values of the
growth rates in the equatorial models are about 5 times greater, or more, than
the growth rates in the polar models. Therefore, excitation in the
composite model requires that envelope convection is suppressed at nearly all
latitudes, except for the case of $l=1$, $m=0$ modes, which become
unstable if  $\vartheta$ is greater than about 60$^\circ$.  The suppression of
envelope convection in a rather large angular region could be a
consequence of the intensity of the magnetic field present in
$\gamma$~Equ, whose mean field modulus was estimated to be  B$\sim$~4.0kG \citep{ryabchikova97}. However, there is little knowledge about
the extent to which a magnetic field can
suppress convection once it becomes nearly horizontal, as would
happen in the case of a dipolar magnetic field close to the magnetic
equator \citep{goughandtayler66,mossandtayler69}.  Hence, the conjecture that suppression
of convection may extend to such large co-latitudes goes without proof.

Finally, we consider the case of $\alpha$~Cir. Unlike the other
targets, the observations of this star raise a clear problem to the driving
model discussed in this work. Inspection of the envelope of positive
growth rates in polar models of $\alpha$~Cir would lead us to the
conclusion that, if high frequency oscillations were to be excited in
this star, their frequencies should be around 1.7~mHz, rather than
around 2.4~mHz, where they are observed.  
 We note, however, that it is not the high value of the frequencies 
 of the oscillations by itself that prevents the agreement of the
 model with the observations. In fact, according to figure 4 of
 \cite{cunha02}, polar models can predict unstable
 modes with frequencies as high as 3.2~mHz, in stellar models of
 relatively low luminosity and close to the Zero Age Main Sequence.
 Thus, it is the combination of high frequency oscillations and
 relatively large stellar radius (of $\sim$1.97~R$_{\odot}$) that makes it
 impossible for model results to agree with the observations in the case of
 $\alpha$~Cir.

\section{Discussion}
\label{discussion}
The results presented in the preceding section highlight the
importance of having access to accurate determinations of global
parameters of stellar pulsators in order to test theoretical models such as the
non-adiabatic models discussed here. In the case of roAp stars that is
even more evident, because their chemical peculiar nature makes the
determination of the effective temperature particularly complex. 
Considering the four roAp stars for which angular diameters
determined based on interferometric data are currently available, we
were able to verify that for three of them the frequency region in which modes are
predicted to be unstable, according to the models described in
\cite{balmforthetal01} and \cite{cunha02}, agrees well with the region
where frequencies are observed. Based on the comparison of model
results and observations we were also able to put constraints on the minimum extent of the angular region in which envelope
convection must be suppressed, according to the proposed models.

On the other hand, we found that the fourth star in our sample, namely $\alpha$~Cir,
represents a challenge to the excitation models proposed by
\cite{balmforthetal01}. One aspect of potential importance in this study
is how the acoustic cutoff frequency in the models compares with the observed
frequencies. As discussed in Section \ref{models}, one of the caveats
of the models considered here is that they do not account for the
direct effect of the magnetic field on pulsations. Oscillations with
frequency above the acoustic cutoff are particularly sensitive to the
real dynamics of  the atmospheric layers of the star, where the
magnetic field plays an important role. Because they neglect that effect, our
models become particularly inadequate to study these high frequency
oscillations. Inspection of the default models for the four stars in
our sample leads us to the following values for their acoustic cutoff
frequencies: $\nu_c\sim 1.6$~mHz for $\gamma$~Equ; $\nu_c\sim 1.5$~mHz
for 10~Aql; $\nu_c\sim 1.1$~mHz for  $\beta$~CrB; $\nu_c\sim 1.9$~mHz
for $\alpha$~Cir.  Clearly, in the first three stars the observed
oscillations are below (albeit close) to the critical cutoff
frequency, while in the case of $\alpha$~Cir they are well above the
latter.  This could be part of the reason for our lack of success
with $\alpha$~Cir models.  However, the fact that the frequency range
of excitation scales with the radius of the model in a well defined manner,
together with the fact that the observed frequencies in this star are
not placed where the scaling would predict,
must lead us to the conclusion that an intrinsic
difference exists between the case of  $\alpha$~Cir and the case of
the other
three stars. 

Among the roAp stars discovered so far we could find another 11 cases,
in addition to $\alpha$~Cir, that oscillate above $\sim$~2~mHz. Unfortunately, a fraction of these
do not have a precise parallax determination, and among those that do, only two have
radii estimated based on advanced atmospheric models
such as those used in the derivation of the spectroscopic inputs
considered in this work. Hence,
only for these two targets a reasonable test of agreement between
model predictions and observations may be attempted (we remind the
reader that polar models predict unstable oscillations as high as
3.2~mHz, thus the fact that frequencies are above 2~mHz, by itself,
does not constitute 
a problem to the theory). These stars are
HD 24712 (HR 1217), which pulsates in the frequency range
$\sim$2.6~--~2.8~mHz \citep{kurtzetal05}, and HD 137949 (33 Lib),
with pulsations in the range
$\sim$1.8~--~2.0~mHz \citep{sachkovetal11}. Their
radii, as derived from advanced atmospheric models, are, respectively, 
$R$=1.778~R$_\odot$ for HD 24712 \citep{shulyaketal2009} and
$R$=2.13~R$_\odot$ for HD 137949 \citep{shulyak2013}.

Accepting the global parameters derived by
\cite{shulyaketal2009,shulyak2013},  we find that HD 24712 is slightly
beyond the red
edge of the instability strip derived by \cite{cunha02}, meaning that in default polar
models of this star no high frequencies are expected to be found unstable. 
If we ignore this fact for a moment and consider the scaling with radius
of the envelop of growth rates (ignoring the small mass dependence
expected in that scaling), we find, in addition, that the envelope would be centered 
around 2.0~mHz, for the radius considered, thus significantly out of
place in relation to the observed frequencies.  For these two reasons
it is reasonable to accept that HD 24712 is an example of the group of roAp
stars whose pulsations cannot be explained by the excitation in models
with convection suppressed.
On the other hand, HD 137949 is located well within the instability
strip predicted by default polar models, but if we consider the same
scaling we find that its radius would have to be $\sim$ 1.8~R$_\odot$ in order
for its pulsations to be explained by the models under discussion. So,
either this star is significantly hotter than that predicted by
\cite{shulyak2013}, with an effective temperature around 8000~K, or,
more likely, it is another example of the same group.

   \begin{figure}
  \centering
   \includegraphics[width=7.cm]{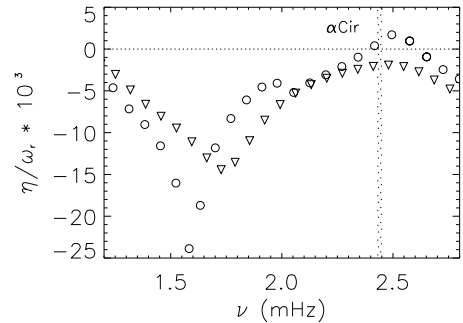}
      \caption{Normalized growth rates in equatorial models with
        different atmospheric properties. Both models were computed
        with photometric inputs. Open circles show results for a
        model similar to the default equatorial model, but with helium settling and
        $Y_{\rm surf}$=0.08. Open triangles are for a model with a
        solar-like atmosphere in which the $T$-$\tau$ relation is
        derived from a model C of Vernazza et al.\, adopting
        $\tau_{\rm min}=10^{-4}$. 
              }
         \label{exotic}
   \end{figure}

In summary, if we extend our sample to include also HD~24712 and HD~137949, for which there are no interferometric data, but there are
radii determinations based on advanced stellar models that account for
element stratification,  we find that in the new
sample of 6 stars, half exhibit oscillations in the
frequency region predicted by the models with convection suppressed,
which is close, but below the acoustic cutoff frequency of the models,
while in the other half, the observed oscillations are significantly
above the predicted region. Thus there is a group of roAp stars,
namely that in which observed pulsation frequencies are clearly above the critical cutoff
frequency expected for their position in the HR diagramme, for which
the driving of the oscillations cannot currently be explained by the models 
proposed by \cite{balmforthetal01}.  We should add that this
conclusion can hardly be changed by varying the inputs to the
equilibrium model, such as the interior metallicity or helium abundance. 
The reason is that such changes,
while keeping the radius fixed, will impact on the results mostly
through a small change in the mass of the model, which will hardly
influence the position of the envelope of excited modes. 
 Given that the
driving in the alternative non-adiabatic models proposed in the
literature \citep{gautschy98,saio05,theadoetal09} originates also on the opacity mechanism working on the
hydrogen ionization region,  we expect that similar difficulties
would be found when attempting to use these models to explain the
observations (and in fact, such was confirmed by H. Saio in a private
communication).  

It is worth noting that while studying equatorial models
for $\alpha$~Cir with different atmospheres or with
He diffusion, we succeeded in finding cases in which modes similar to
those observed in this star are excited, or in which their growth rates, while still negative,
approach zero. An example of the results for such models, namely, a
model with a solar-like atmosphere in which the $T$-$\tau$ relation is
derived from a model C of \cite{vernazzaetal81} and a model  similar
to the default equatorial model, but with helium settling and $Y_{\rm
  surf}$=0.08, is shown in Fig. \ref{exotic}. Inspection of the work integrals shows
that in these cases the strongest contribution for the mode energy
input still comes from the region of hydrogen ionization, but in this
case the driving agent is the turbulent pressure, rather than the opacity.
 In fact, a similar driving effect of the turbulent pressure has been seen also in $\delta$-Scuti
stars by  Antoci et al. (in preparation)  for intermediate radial order modes. 
Unfortunately, since our models do not retain the complete dynamics
experienced by these very high
frequency oscillations in the atmospheric layers, we refrain from
further exploring this potential solution for the driving of the
oscillations in stars like $\alpha$~Cir. Nevertheless, these results indicate that excitation of
modes with frequencies above the acoustic cutoff  in the hydrogen ionization
region is not impossible, in models with convection.

\section{acknowledgements}
MSC is supported by an Investigador FCT contract funded by
FCT/MCTES (Portugal) and POPH/FSE (EC).  IMB acknowledges financial support from the FCT (Portugal) through the
SFRH/BPD/87857/2012. This research made
use of funds from the ERC, under FP7/EC, through the project
FP7-SPACE-2012-312844. Part of this research was funded through an International
Research Collaboration Award of the University of Sydney.

\bibliography{roap} 

\begin{thebibliography}{45}
\expandafter\ifx\csname natexlab\endcsname\relax\def\natexlab#1{#1}\fi

\bibitem[{{Alentiev} {et~al.}(2012){Alentiev}, {Kochukhov}, {Ryabchikova},
  {Cunha}, {Tsymbal}, \& {Weiss}}]{alentievetal12}
{Alentiev} D., {Kochukhov} O., {Ryabchikova} T., {Cunha} M., {Tsymbal} V.,
  {Weiss} W., 2012, \mnras, 421, L82

\bibitem[{{Asplund} {et~al.}(2009){Asplund}, {Grevesse}, {Sauval}, \&
  {Scott}}]{Asplund2009}
{Asplund} M., {Grevesse} N., {Sauval} A.~J., {Scott} P., 2009, \araa, 47, 481

\bibitem[{{Balmforth}(1992)}]{balmforth92}
{Balmforth} N.~J., 1992, \mnras, 255, 603

\bibitem[{{Balmforth} {et~al.}(2001){Balmforth}, {Cunha}, {Dolez}, {Gough}, \&
  {Vauclair}}]{balmforthetal01}
{Balmforth} N.~J., {Cunha} M.~S., {Dolez} N., {Gough} D.~O., {Vauclair} S.,
  2001, \mnras, 323, 362

\bibitem[{{Balona} {et~al.}(2011{\natexlab{a}}){Balona}, {Cunha},
  {Gruberbauer}, {Kurtz}, {Saio}, {White}, {Christensen-Dalsgaard}, {Kjeldsen},
  {Christiansen}, {Hall}, \& {Seader}}]{balonaetal11b}
{Balona} L.~A., {Cunha} M.~S., {Gruberbauer} M., {Kurtz} D.~W., {Saio} H.,
  {White} T.~R., {Christensen-Dalsgaard} J., {Kjeldsen} H., {Christiansen}
  J.~L., {Hall} J.~R., {Seader} S.~E., 2011{\natexlab{a}}, \mnras, 413, 2651

\bibitem[{{Balona} {et~al.}(2011{\natexlab{b}}){Balona}, {Cunha}, {Kurtz},
  {Brand{\~a}o}, {Gruberbauer}, {Saio}, {{\"O}stensen}, {Elkin}, {Borucki},
  {Christensen-Dalsgaard}, {Kjeldsen}, {Koch}, \& {Bryson}}]{balonaetal11a}
{Balona} L.~A., {Cunha} M.~S., {Kurtz} D.~W., {Brand{\~a}o} I.~M.,
  {Gruberbauer} M., {Saio} H., {{\"O}stensen} R., {Elkin} V.~G., {Borucki}
  W.~J., {Christensen-Dalsgaard} J., {Kjeldsen} H., {Koch} D.~G., {Bryson}
  S.~T., 2011{\natexlab{b}}, \mnras, 410, 517

\bibitem[{{Bruntt} {et~al.}(2010){Bruntt}, {Kervella}, {M{\'e}rand},
  {Brand{\~a}o}, {Bedding}, {Ten Brummelaar}, {Coud{\'e} Du Foresto}, {Cunha},
  {Farrington}, {Goldfinger}, {Kiss}, {McAlister}, {Ridgway}, {Sturmann},
  {Sturmann}, {Turner}, \& {Tuthill}}]{brunttetal10}
{Bruntt} H., {Kervella} P., {M{\'e}rand} A., {Brand{\~a}o} I.~M., {Bedding}
  T.~R., {Ten Brummelaar} T.~A., {Coud{\'e} Du Foresto} V., {Cunha} M.~S.,
  {Farrington} C., {Goldfinger} P.~J., {Kiss} L.~L., {McAlister} H.~A.,
  {Ridgway} S.~T., {Sturmann} J., {Sturmann} L., {Turner} N., {Tuthill} P.~G.,
  2010, \aap, 512, A55

\bibitem[{{Bruntt} {et~al.}(2009){Bruntt}, {Kurtz}, {Cunha}, {Brand{\~a}o},
  {Handler}, {Bedding}, {Medupe}, {Buzasi}, {Mashigo}, {Zhang}, \& {van
  Wyk}}]{brunttetal09}
{Bruntt} H., {Kurtz} D.~W., {Cunha} M.~S., {Brand{\~a}o} I.~M., {Handler} G.,
  {Bedding} T.~R., {Medupe} T., {Buzasi} D.~L., {Mashigo} D., {Zhang} I., {van
  Wyk} F., 2009, \mnras, 396, 1189

\bibitem[{{Bruntt} {et~al.}(2008){Bruntt}, {North}, {Cunha}, {Brand{\~a}o},
  {Elkin}, {Kurtz}, {Davis}, {Bedding}, {Jacob}, {Owens}, {Robertson}, {Tango},
  {Gameiro}, {Ireland}, \& {Tuthill}}]{brunttetal08}
{Bruntt} H., {North} J.~R., {Cunha} M., {Brand{\~a}o} I.~M., {Elkin} V.~G.,
  {Kurtz} D.~W., {Davis} J., {Bedding} T.~R., {Jacob} A.~P., {Owens} S.~M.,
  {Robertson} J.~G., {Tango} W.~J., {Gameiro} J.~F., {Ireland} M.~J., {Tuthill}
  P.~G., 2008, \mnras, 386, 2039

\bibitem[{{Cunha}(2002)}]{cunha02}
{Cunha} M.~S., 2002, \mnras, 333, 47

\bibitem[{{Cunha}(2007)}]{cunha07}
---, 2007, Communications in Asteroseismology, 150, 48

\bibitem[{{Cunha} {et~al.}(2007){Cunha}, {Aerts}, {Christensen-Dalsgaard},
  {Baglin}, {Bigot}, {Brown}, {Catala}, {Creevey}, {de Souza}, {Eggenberger},
  {Garcia}, {Grundahl}, {Kervella}, {Kurtz}, {Mathias}, {Miglio}, {Monteiro},
  {Perrin}, {Pijpers}, {Pourbaix}, {Quirrenbach}, {Rousselet-Perraut},
  {Teixeira}, {Th{\'e}venin}, \& {Thompson}}]{cunhaetal07}
{Cunha} M.~S., {Aerts} C., {Christensen-Dalsgaard} J., {Baglin} A., {Bigot} L.,
  {Brown} T.~M., {Catala} C., {Creevey} O.~L., {de Souza} A.~D., {Eggenberger}
  P., {Garcia} P.~J.~V., {Grundahl} F., {Kervella} P., {Kurtz} D.~W., {Mathias}
  P., {Miglio} A., {Monteiro} M.~J.~P.~F.~G., {Perrin} G., {Pijpers} F.~P.,
  {Pourbaix} D., {Quirrenbach} A., {Rousselet-Perraut} K., {Teixeira} T.~C.,
  {Th{\'e}venin} F., {Thompson} M.~J., 2007, Astronomy and Astrophysics Review,
  14, 217

\bibitem[{{Elkin} {et~al.}(2005){Elkin}, {Riley}, {Cunha}, {Kurtz}, \&
  {Mathys}}]{elkinetal05}
{Elkin} V.~G., {Riley} J.~D., {Cunha} M.~S., {Kurtz} D.~W., {Mathys} G., 2005,
  \mnras, 358, 665

\bibitem[{{Gautschy} {et~al.}(1998){Gautschy}, {Saio}, \&
  {Harzenmoser}}]{gautschy98}
{Gautschy} A., {Saio} H., {Harzenmoser} H., 1998, \mnras, 301, 31

\bibitem[{{Gough}(1977)}]{gough77}
{Gough} D.~O., 1977, in Lecture Notes in Physics, Berlin Springer Verlag,
  Vol.~71, Problems of Stellar Convection, {Spiegel} E.~A., {Zahn} J.-P., eds.,
  pp. 349--363

\bibitem[{{Gough} \& {Tayler}(1966)}]{goughandtayler66}
{Gough} D.~O., {Tayler} R.~J., 1966, \mnras, 133, 85

\bibitem[{{Grevesse} \& {Noels}(1993)}]{grevesseandnoels93}
{Grevesse} N., {Noels} A., 1993, in Origin and Evolution of the Elements,
  {Prantzos} N., {Vangioni-Flam} E., {Casse} M., eds., pp. 15--25

\bibitem[{{Gruberbauer} {et~al.}(2008){Gruberbauer}, {Saio}, {Huber},
  {Kallinger}, {Weiss}, {Guenther}, {Kuschnig}, {Matthews}, {Moffat},
  {Rucinski}, {Sasselov}, \& {Walker}}]{gruberbaueretal08}
{Gruberbauer} M., {Saio} H., {Huber} D., {Kallinger} T., {Weiss} W.~W.,
  {Guenther} D.~B., {Kuschnig} R., {Matthews} J.~M., {Moffat} A.~F.~J.,
  {Rucinski} S., {Sasselov} D., {Walker} G.~A.~H., 2008, \aap, 480, 223

\bibitem[{{Hatzes} \& {Mkrtichian}(2004)}]{hatzes04}
{Hatzes} A.~P., {Mkrtichian} D.~E., 2004, \mnras, 351, 663

\bibitem[{{Huber} {et~al.}(2008){Huber}, {Saio}, {Gruberbauer}, {Weiss},
  {Rowe}, {Hareter}, {Kallinger}, {Reegen}, {Matthews}, {Kuschnig}, {Guenther},
  {Moffat}, {Rucinski}, {Sasselov}, \& {Walker}}]{huberetal08}
{Huber} D., {Saio} H., {Gruberbauer} M., {Weiss} W.~W., {Rowe} J.~F., {Hareter}
  M., {Kallinger} T., {Reegen} P., {Matthews} J.~M., {Kuschnig} R., {Guenther}
  D.~B., {Moffat} A.~F.~J., {Rucinski} S., {Sasselov} D., {Walker} G.~A.~H.,
  2008, \aap, 483, 239

\bibitem[{{Hubrig} {et~al.}(2005){Hubrig}, {Nesvacil}, {Sch{\"o}ller}, {North},
  {Mathys}, {Kurtz}, {Wolff}, {Szeifert}, {Cunha}, \& {Elkin}}]{hubrigetal05}
{Hubrig} S., {Nesvacil} N., {Sch{\"o}ller} M., {North} P., {Mathys} G., {Kurtz}
  D.~W., {Wolff} B., {Szeifert} T., {Cunha} M.~S., {Elkin} V.~G., 2005, \aap,
  440, L37

\bibitem[{{Kochukhov}(2009)}]{2009CoAst.159...61K}
{Kochukhov} O., 2009, Communications in Asteroseismology, 159, 61

\bibitem[{{Kochukhov} \& {Bagnulo}(2006)}]{Kochukhovandbagnulo06}
{Kochukhov} O., {Bagnulo} S., 2006, \aap, 450, 763

\bibitem[{{Kochukhov} {et~al.}(2009){Kochukhov}, {Shulyak}, \&
  {Ryabchikova}}]{kochukhov2009}
{Kochukhov} O., {Shulyak} D., {Ryabchikova} T., 2009, \aap, 499, 851

\bibitem[{{Kreidl} \& {Kurtz}(1986)}]{kreidl86}
{Kreidl} T.~J., {Kurtz} D.~W., 1986, \mnras, 220, 313

\bibitem[{{Kurtz}(1982)}]{kurtz82}
{Kurtz} D.~W., 1982, \mnras, 200, 807

\bibitem[{{Kurtz} {et~al.}(2005){Kurtz}, {Cameron}, {Cunha}, {Dolez},
  {Vauclair}, {Pallier}, {Ulla}, {Kepler}, {da Costa}, {Kanaan}, {Fraga},
  {Giovannini}, {Wood}, {Silvestri}, {Kawaler}, {Riddle}, {Reed}, {Watson},
  {Metcalfe}, {Mukadam}, {Nather}, {Winget}, {Nitta}, {Kleinman}, {Guzik},
  {Bradley}, {Matthews}, {Sekiguchi}, {Sullivan}, {Sullivan}, {Shobbrook},
  {Jiang}, {Birch}, {Ashoka}, {Seetha}, {Girish}, {Joshi}, {Moskalik}, {Zola},
  {O'Donoghue}, {Handler}, {Mueller}, {Gonzalez Perez}, {Solheim},
  {Johannessen}, \& {Bigot}}]{kurtzetal05}
{Kurtz} D.~W., {Cameron} C., {Cunha} M.~S., {Dolez} N., {Vauclair} G.,
  {Pallier} E., {Ulla} A., {Kepler} S.~O., {da Costa} A., {Kanaan} A., {Fraga}
  L., {Giovannini} O., {Wood} M.~A., {Silvestri} N., {Kawaler} S.~D., {Riddle}
  R.~L., {Reed} M.~D., {Watson} T.~K., {Metcalfe} T.~S., {Mukadam} A., {Nather}
  R.~E., {Winget} D.~E., {Nitta} A., {Kleinman} S.~J., {Guzik} J.~A., {Bradley}
  P.~A., {Matthews} J.~M., {Sekiguchi} K., {Sullivan} D.~J., {Sullivan} T.,
  {Shobbrook} R., {Jiang} X., {Birch} P.~V., {Ashoka} B.~N., {Seetha} S.,
  {Girish} V., {Joshi} S., {Moskalik} P., {Zola} S., {O'Donoghue} D., {Handler}
  G., {Mueller} M., {Gonzalez Perez} J.~M., {Solheim} J.-E., {Johannessen} F.,
  {Bigot} L., 2005, \mnras, 358, 651

\bibitem[{{Kurtz} {et~al.}(2011){Kurtz}, {Cunha}, {Saio}, {Bigot}, {Balona},
  {Elkin}, {Shibahashi}, {Brand{\~a}o}, {Uytterhoeven}, {Frandsen}, {Frimann},
  {Hatzes}, {Lueftinger}, {Gruberbauer}, {Kjeldsen}, {Christensen-Dalsgaard},
  \& {Kawaler}}]{kurtzetal11}
{Kurtz} D.~W., {Cunha} M.~S., {Saio} H., {Bigot} L., {Balona} L.~A., {Elkin}
  V.~G., {Shibahashi} H., {Brand{\~a}o} I.~M., {Uytterhoeven} K., {Frandsen}
  S., {Frimann} S., {Hatzes} A., {Lueftinger} T., {Gruberbauer} M., {Kjeldsen}
  H., {Christensen-Dalsgaard} J., {Kawaler} S.~D., 2011, \mnras, 414, 2550

\bibitem[{{Kurtz} {et~al.}(2007){Kurtz}, {Elkin}, \& {Mathys}}]{kurtzetal07}
{Kurtz} D.~W., {Elkin} V.~G., {Mathys} G., 2007, \mnras, 380, 741

\bibitem[{{Moss} \& {Tayler}(1969)}]{mossandtayler69}
{Moss} D.~L., {Tayler} R.~J., 1969, \mnras, 145, 217

\bibitem[{{Nesvacil} {et~al.}(2013){Nesvacil}, {Shulyak}, {Ryabchikova},
  {Kochukhov}, {Akberov}, \& {Weiss}}]{nesvacil2013}
{Nesvacil} N., {Shulyak} D., {Ryabchikova} T.~A., {Kochukhov} O., {Akberov} A.,
  {Weiss} W., 2013, \aap, 552, A28

\bibitem[{{Perraut} {et~al.}(2011){Perraut}, {Brand{\~a}o}, {Mourard}, {Cunha},
  {B{\'e}rio}, {Bonneau}, {Chesneau}, {Clausse}, {Delaa}, {Marcotto},
  {Roussel}, {Spang}, {Stee}, {Tallon-Bosc}, {McAlister}, {Ten Brummelaar},
  {Sturmann}, {Sturmann}, {Turner}, {Farrington}, \&
  {Goldfinger}}]{perrautetal11}
{Perraut} K., {Brand{\~a}o} I., {Mourard} D., {Cunha} M., {B{\'e}rio} P.,
  {Bonneau} D., {Chesneau} O., {Clausse} J.~M., {Delaa} O., {Marcotto} A.,
  {Roussel} A., {Spang} A., {Stee} P., {Tallon-Bosc} I., {McAlister} H., {Ten
  Brummelaar} T., {Sturmann} J., {Sturmann} L., {Turner} N., {Farrington} C.,
  {Goldfinger} P.~J., 2011, \aap, 526, A89

\bibitem[{{Ryabchikova} {et~al.}(1997){Ryabchikova}, {Adelman}, {Weiss}, \&
  {Kuschnig}}]{ryabchikova97}
{Ryabchikova} T.~A., {Adelman} S.~J., {Weiss} W.~W., {Kuschnig} R., 1997, \aap,
  322, 234

\bibitem[{{Sachkov} {et~al.}(2011){Sachkov}, {Hareter}, {Ryabchikova}, {Wade},
  {Kochukhov}, {Shulyak}, \& {Weiss}}]{sachkovetal11}
{Sachkov} M., {Hareter} M., {Ryabchikova} T., {Wade} G., {Kochukhov} O.,
  {Shulyak} D., {Weiss} W.~W., 2011, \mnras, 416, 2669

\bibitem[{{Saio}(2005)}]{saio05}
{Saio} H., 2005, \mnras, 360, 1022

\bibitem[{{Saio} {et~al.}(2012){Saio}, {Gruberbauer}, {Weiss}, {Matthews}, \&
  {Ryabchikova}}]{saioetal12}
{Saio} H., {Gruberbauer} M., {Weiss} W.~W., {Matthews} J.~M., {Ryabchikova} T.,
  2012, \mnras, 420, 283

\bibitem[{{Serenelli} \& {Basu}(2010)}]{serenelli2010}
{Serenelli} A.~M., {Basu} S., 2010, \apj, 719, 865

\bibitem[{{Shibahashi} \& {Saio}(1985)}]{shibahashi85}
{Shibahashi} H., {Saio} H., 1985, \pasj, 37, 245

\bibitem[{{Shulyak} {et~al.}(2013){Shulyak}, {Ryabchikova}, \&
  {Kochukhov}}]{shulyak2013}
{Shulyak} D., {Ryabchikova} T., {Kochukhov} O., 2013, \aap, 551, A14

\bibitem[{{Shulyak} {et~al.}(2009){Shulyak}, {Ryabchikova}, {Mashonkina}, \&
  {Kochukhov}}]{shulyaketal2009}
{Shulyak} D., {Ryabchikova} T., {Mashonkina} L., {Kochukhov} O., 2009, \aap,
  499, 879

\bibitem[{{Shulyak} {et~al.}(2004){Shulyak}, {Tsymbal}, {Ryabchikova},
  {St{\"u}tz}, \& {Weiss}}]{shulyak2004}
{Shulyak} D., {Tsymbal} V., {Ryabchikova} T., {St{\"u}tz} C., {Weiss} W.~W.,
  2004, \aap, 428, 993

\bibitem[{{Sousa} \& {Cunha}(2008)}]{sousa08}
{Sousa} S.~G., {Cunha} M.~S., 2008, \mnras, 386, 531

\bibitem[{{Th{\'e}ado} {et~al.}(2009){Th{\'e}ado}, {Dupret}, {Noels}, \&
  {Ferguson}}]{theadoetal09}
{Th{\'e}ado} S., {Dupret} M.-A., {Noels} A., {Ferguson} J.~W., 2009, \aap, 493,
  159

\bibitem[{{Th{\'e}ado} {et~al.}(2005){Th{\'e}ado}, {Vauclair}, \&
  {Cunha}}]{theadoetal05}
{Th{\'e}ado} S., {Vauclair} S., {Cunha} M.~S., 2005, \aap, 443, 627

\bibitem[{{Vernazza} {et~al.}(1981){Vernazza}, {Avrett}, \&
  {Loeser}}]{vernazzaetal81}
{Vernazza} J.~E., {Avrett} E.~H., {Loeser} R., 1981, \apjs, 45, 635

\end{thebibliography}

\end{document}